\documentclass[aps,prl,twocolumn,groupedaddress,floatfix,showpacs,showkeys]{revtex4}

\usepackage[utf8]{inputenc}
\usepackage[T1]{fontenc}
\usepackage{slashed}
\usepackage{times}

\usepackage{amssymb,amsfonts,amsmath,amsthm}
\usepackage{dsfont,bbm}
\usepackage{dcolumn}

\usepackage{graphicx}
\usepackage{color}

\newcommand{\intp}{\mathcal{I}}
\newcommand{\braket}[1]{\langle #1 \rangle}

\begin{document}

\title{Expanding the Interpolator Basis in the Variational Method to Explicitly Account for Backward Running States}

\author{Rainer W.~{Schiel}}
   \email{rainer.schiel@ur.de}
   \affiliation{Institut f\"ur Theoretische Physik, Universit\"at Regensburg, 93040 Regensburg, Germany}

\date{\today}

\begin{abstract}
In this paper, I show that backward (in time) running states can be explicitly accounted for by expanding the interpolator basis in the variational method in lattice QCD. 
The backward running states can then be removed by choosing an appropriate linear combination of interpolators, which improves the signal significantly.
The proof of principle, which also makes use of the Time-Shift Trick (Generalized Pencil-of-Functions method), will be delivered at an example on a $64^4$ lattice close to the physical pion mass.
\end{abstract}


\maketitle

\paragraph{Introduction.---}

In lattice QCD, the variational method \cite{Michael:1982gb, Michael:1985ne, Luscher:1990ck, Blossier:2009kd} has become a widely used method to separate physical states. 
Its main application is spectroscopy (recent examples are, e.g., \cite{Lang:2012db, Edwards:2012fx, Alexandrou:2013fsu, Bali:2013kia, Dudek:2013yja, Mahbub:2013bba, Lang:2014tia, Denissenya:2014ywa, Detar:2014gla, Perez:2014jra}), but it has also been successfully used to extract specific physical states in hadron structure calculations (e.g., \cite{Owen:2012ts, Menadue:2013xqa, Owen:2013pfa, Braun:2014wpa, Mastropas:2014fsa}).
In this paper, I show that backward (in time) running states can be explicitly accounted for by expanding the interpolator basis in the variational method. 
This is important for studies where the backward propagating states are not negligible compared to the forward propagating ones in the time range of interest.
The backward propagating states can then be eliminated with a properly chosen linear combination of interpolators.
For the proof of principle, the additional interpolators are obtained ``for free'' using the Time-Shift Trick (Generalized Pencil-of-Function method), which I will also briefly review in this paper.

\paragraph{The Variational Method.---}

\begin{figure}
    \includegraphics[width=\unitlength]{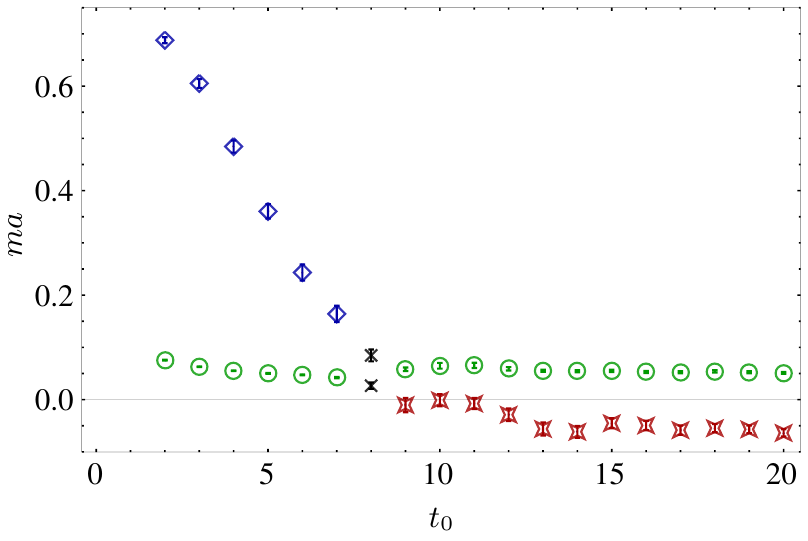}%
\caption{The mass eigenvalues, found with the traditional variational method with two interpolators, applied to the pion. 
For $t \geq 9$, this method extracts the forward (green circles) and backward (red stars) running pion.
For $t \leq 7$, it finds the pion and an excited state (blue diamonds), but the excited state does not form a plateau.
For $t = 8$, the identification of the found states is unclear (black crosses).} \label{fig:varmeth22}
\end{figure}

The standard variational method uses several interpolators $ \intp_i , i = 1, \dots, N_\intp$ to separate the physical states $ s_i , i = 1, \dots, N_s$, with energies $E_i$. 
Typically, one assumes that the number of significantly contributing physical states is equal to the number of interpolators $N_\intp = N_s \equiv N$. 
The interpolators couple to the physical states through
\[
\braket{s_i | \intp_j^\dagger | 0} = a_{ij}.
\]
On a lattice with (anti)periodic boundary conditions, the correlation matrix is
\begin{align} \label{eq:corrmat}
C_{ij} (t) \equiv &\, \braket{\intp_i (t) | \intp_j^\dagger (0)} \nonumber \\
=&\, \sum_k \braket{0 | \intp_i | s_k} \braket{s_k | \intp_j^\dagger | 0} \exp(-E_k t) \nonumber \\
&\, + \sum_k \braket{0 | \intp_j^\dagger | s_k} \braket{s_k | \intp_i | 0} \exp(-E_k (T-t)) \nonumber \\
=&\, \sum_k a^\dagger_{ik} a_{kj} \exp( -E_k t) \nonumber \\
&\, + \sum_k b^\dagger_{ik} b_{kj} \exp(-E_k (T - t)),
\end{align}
where $T$ is the temporal extent of the lattice and I have introduced $b_{ij} \equiv \braket{0 | \intp^\dagger_j | s_i}$.
The first term in Eq.~\eqref{eq:corrmat} contains the forward moving and the second term the backward moving states.
Neglecting the backward moving states (which is often a good approximation), Eq.~\eqref{eq:corrmat} becomes
\[ 
C_{ij} (t) = \sum_k a^\dagger_{ik} a_{kj} \exp( -E_k t). 
\]
One proceeds by solving the generalized eigenvalue problem
\[
\sum_j C_{ij}(t_1) v^{(k)}_j = \lambda^{(k)} \sum_j C_{ij}(t_0) v^{(k)}_j,
\]
where I denote the $k^\text{th}$ eigenvalue and -vector with $\lambda^{(k)}$ and $v^{(k)}_j$, respectively.
This is equivalent to finding the eigenvectors and -values of the matrix 
\begin{align}
G_{ij}(t_0, t_1) &\equiv\, \sum_k C_{ik}^{-1}(t_0) C_{kj}(t_1) \label{eq:varmatrix} \\
&=\, \sum_k a_{ik}^{-1} a_{kj} \exp(-E_k (t_1 - t_0)). \nonumber
\end{align}
One sees immediately that the eigenvalues are $\exp(-E_i (t_1 - t_0))$ and the $i^\text{th}$ eigenvector, with components $j$, is $v^{(i)}_j = a_{ji}^{-1}$.
We can then construct optimal interpolators $\intp_\text{opt}^{(i) \dagger} \equiv \sum_j \intp^\dagger_j v^{(i)}_j$ which couple -- at least in principle -- only to the physical state $s_i$:
\[
\braket{s_k | \intp^{(i) \dagger}_\text{opt} | 0} = \sum_j \braket{s_k | \intp^\dagger_j | 0} v_j^{(i)} = \sum_j a_{kj} a^{-1}_{ji} = \delta_{ki}.
\]
While this method works extremely well in many cases (see, e.g., \cite{Lang:2012db, Edwards:2012fx, Alexandrou:2013fsu, Bali:2013kia, Dudek:2013yja, Mahbub:2013bba, Lang:2014tia, Denissenya:2014ywa, Detar:2014gla, Perez:2014jra, Owen:2012ts, Menadue:2013xqa, Owen:2013pfa, Braun:2014wpa, Mastropas:2014fsa}), it fails for lattices with (anti)periodic boundary conditions when the backward running states are so light that they do not decay sufficiently fast and contribute significantly on the left side.
In Figure \ref{fig:varmeth22}, I show this problem for pions on a $64^4$ ensemble with Wilson gauge action, $N_f = 2$ flavors of dynamical Wilson (Clover) fermions, $\beta = 5.29$ (corresponding to $a \approx 0.07\ \text{fm}$) and $\kappa = 0.13640$, where the pion has a mass of approximately $150\ \text{MeV}$.
For ensembles like this one, but also at much larger pion masses, the variational method has to be modified to yield reasonable results.

\setlength{\unitlength}{0.48\textwidth}

\paragraph{The $\cosh$ Method.---}

The straightforward way to account for the backward running states is to replace the exponential decays by $\cosh$s (and $\sinh$s, see below). 
As an example, I use two interpolators for the pion ($\pi$) and excited pion ($\pi^*$) states, $\intp_1 = \Pi = \bar{d} \gamma_5 u$ and $\intp_2 = A_0 = \bar{d} \gamma_5 \gamma_4 u$.
The adjoint interpolators, which couple to the antiparticle, are: 
\begin{align}
\Pi^\dagger &=\, \left( \bar{d} \gamma_5 u \right)^\dagger = \bar{u} \gamma_5 d \nonumber \\
A_0^\dagger &=\, \left( \bar{d} \gamma_5 \gamma_4 u \right)^\dagger = \bar{u} \gamma_4 \gamma_5 d = -\bar{u} \gamma_5 \gamma_4 d. \nonumber  
\end{align} 
In simulations in which the up and down quarks are identical (i.e., most of the current simulations), one finds
\begin{align}
\braket{\pi^+ | \Pi^\dagger | 0} &=\, \braket{\pi^- | \Pi | 0} \nonumber \\
\braket{\pi^+ | A_0^\dagger | 0} &=\, - \braket{\pi^- | A_0 | 0}, \nonumber
\end{align}
and equivalently for the excited pion.
In other words, $b^\dagger_{1i} = a_{i1}$ and $b^\dagger_{2i} = - a_{i2}$.
The relative minus sign between these expressions is responsible for the fact that $\braket{A_0 | \Pi^\dagger}$ and $\braket{\Pi | A_0^\dagger}$ have a $\sinh$-like time dependence, while  $\braket{\Pi | \Pi^\dagger}$ and $\braket{A_0 | A_0^\dagger}$ are $\cosh$-like. 
Assuming that the $a_{ij}$ are real, Eq.~\eqref{eq:corrmat} becomes
\begin{align} \label{eqs:cosh}
\braket{\Pi (t) | \Pi^\dagger (0)} =&\, 2 a_{11}^2 \exp(-E_\pi T / 2) \cosh(E_\pi \bar{t}) \nonumber \\
&\,+ 2 a_{21}^2 \exp(-E_* T / 2) \cosh(E_* \bar{t})\nonumber \\
\braket{ A_0 (t) | A_0^\dagger (0) } =&\, 2 a_{12}^2 \exp(-E_\pi T / 2) \cosh(E_\pi \bar{t}) \nonumber \\
&\,+ 2 a_{22}^2 \exp(-E_* T / 2) \cosh(E_* \bar{t})\nonumber \\
\braket{ A_0 (t) | \Pi^\dagger (0) } =&\, -2 a_{11} a_{12} \exp(-E_\pi T / 2) \sinh(E_\pi \bar{t}) \nonumber \\
&\, - 2 a_{21} a_{22} \exp(-E_* T/2) \sinh(E_* \bar{t}) \nonumber \\
=&\, \braket{ \Pi (t) | A_0^\dagger (0) },
\end{align}
where I have used $E_* \equiv E_{\pi^*}$ and $\bar{t} \equiv t - T/2$ for brevity. 
Eqs.~\eqref{eqs:cosh} have to be solved numerically and it turns out that the solution is numerically not very stable and needs some ``supervision''. 
Once the $a_{ij}$ have been found, one can again construct optimal interpolators for the forward running states by $\intp^{(i) \dagger}_\text{opt} = \sum_j \intp^\dagger_j a_{ji}^{-1}$.
One finds for an arbitrary operator $\mathcal{O}$:
\begin{align}
\braket{ \mathcal{O} (t) &| \intp_\text{opt}^{(i) \dagger} (0) } \nonumber \\
=&\, \sum_j \braket{ \mathcal{O} (t) | \intp^\dagger_j (0) } a_{ji}^{-1} \nonumber \\
=&\, \sum_{k, j} \braket{ 0 | \mathcal{O} | s_k } \braket{ s_k | \intp_j^\dagger | 0 } a_{ji}^{-1} \exp(- E_k t) \nonumber \\
&\, + \sum_{k, j} \braket{ 0 | \intp_j^\dagger | s_k } \braket{ s_k | \mathcal{O} | 0 } a_{ji}^{-1} \exp(- E_k (T-t)) \nonumber \\
=&\, \braket{ 0 | \mathcal{O} | s_i } \exp(-E_i t) \nonumber \\
&\, + \sum_{k, j} \braket{ s_k | \mathcal{O} | 0 } b_{kj} a^{-1}_{ji} \exp(-E_k (T-t)), \nonumber
\end{align}
and the example above showed that $b a^{-1}$ is not necessarily the unit matrix.
Therefore, while the forward running contribution is optimized for the wanted state, the backward running part contains several or all states, which prevents simple exponential or $\cosh$ fits.

\paragraph{The Variational Method with an Expanded Interpolator Basis.---}

An improvement over the $\cosh$ method can be achieved by using the standard variational method with an expanded interpolator basis, where the additional interpolators account for the backward running states. 
In other words, I treat the backward running states as forward running particles with negative mass. 
Usually, the heavier backward running states will be negligible on the left side since they decay too fast.
The number of backward running states that contribute significantly is $N_b \leq N_s$. 
One then needs $N_\intp = N_s + N_b$ interpolators to account for the forward and the significant backward running states.
I define $\tilde{b}_{kj} = b_{kj} \exp(-E_k T / 2)$ and sort the states so that the first $N_b$ states are the ones that are non-negligible to rewrite Eq.~\eqref{eq:corrmat} as
\begin{align}
C_{ij}(t) =&\, \sum_{k=1}^{N_s} a^\dagger_{ik} a_{kj} \exp(-E_k t) \nonumber \\
&\, + \sum_{k=1}^{N_b} \tilde{b}^\dagger_{ik} \tilde{b}_{kj} \exp(E_k t). \nonumber
\end{align}
With
\[
d_{kj} \equiv \left\{ \begin{array}{lr@{}l} a_{kj}, & 1 \leq k &\, \leq N_s \\ \tilde{b}_{(k-N_s)j}, & N_s < k &\, \leq N_\intp \end{array}  \right.
\]
and
\[
\tilde{E}_k \equiv \left\{ \begin{array}{lr@{}l} E_k, & 1 \leq k &\, \leq N_s \\ -E_{k-N_s}, & N_s < k &\, \leq N_\intp, \end{array}  \right.
\]
this becomes the standard variational problem. 
I define the optimal interpolator as $\intp_\text{opt}^{(i)\dagger} \equiv \sum_j \intp^\dagger_j d^{-1}_{ji}$.
Then, for an arbitrary operator $\mathcal{O}$, one has (dropping the negligible states)
\begin{align}
\braket{ \mathcal{O} (t) &| \intp^{(i)\dagger}_\text{opt} (0) } \nonumber \\
\approx&\, \sum_{k=1}^{N_s} \sum_j \braket{ 0 | \mathcal{O} | s_k } \braket{ s_k | \intp^\dagger_j | 0 } d^{-1}_{ji} \exp(-E_k t) \nonumber \\
&\, + \sum_{k=1}^{N_b} \sum_j \braket{ 0 | \intp^\dagger_j | s_k } \braket{ s_k | \mathcal{O} | 0 } d^{-1}_{ji} \exp(-E_k (T-t)) \nonumber \\
=&\, \sum_{k=1}^{N_s} \sum_j \braket{ 0 | \mathcal{O} | s_k } a_{kj} d^{-1}_{ji} \exp(-E_k t) \nonumber \\
&\, + \sum_{k=1}^{N_b} \sum_j \braket{ s_k | \mathcal{O} | 0 } \exp(-E_k T/2) \tilde{b}_{kj} d^{-1}_{ji} \exp(E_k t) \nonumber \\
=&\, \left\{ \begin{array}{lr@{}l} 
\braket{ 0 | \mathcal{O} | s_i } \exp(-E_i t), & 1 \leq i & \leq N_s \\
\braket{ s_{i-N_s} | \mathcal{O} | 0 } \exp(E_{i-N_s} \bar{t}), & N_s < i & \leq N_\intp.
\end{array} \right. \nonumber
\end{align}
This shows that the variational method with an expanded interpolator basis works indeed and one is left with an optimal interpolator for the ground or excited states, with no contamination from the backward running states, which greatly improves the further analysis. 

In my example, I need $N_\intp = 3$ interpolators to account for the pion, the excited pion and the backward running pion. 
These interpolators can be obtained using different currents, various levels of smearing or any other method that yields linearly independent interpolators. 
Here, only data for the two interpolators $\Pi$ and $A_0$ with one level of smearing each was available.
So in order to avoid using additional supercomputer time, I had to use the Time-Shift Trick to obtain the third interpolator.

\paragraph{The Time-Shift Trick.---}

\begin{figure}
    \includegraphics[width=\unitlength]{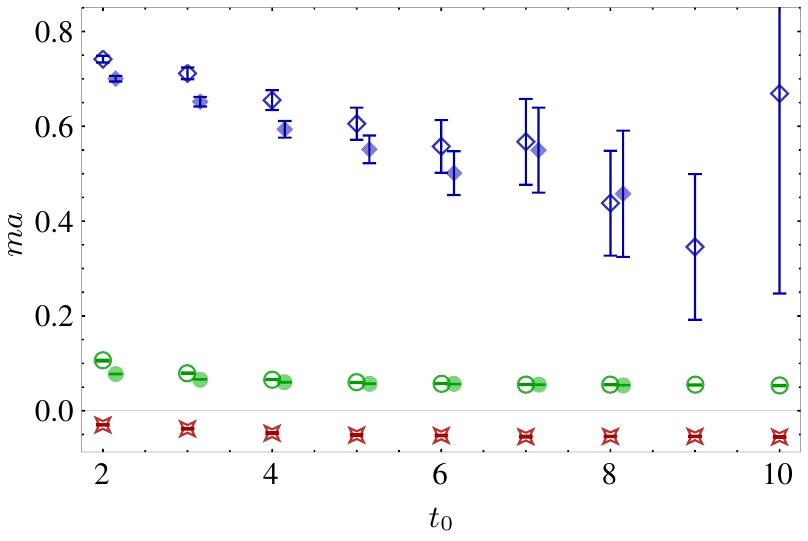}%
\caption{Comparison of the $\cosh$ method (filled symbols, slightly shifted to the right) with the variational method with an expanded interpolator basis (open symbols). 
The backward running pion, the forward running pion and the excited pion state are denoted with red stars, green circles and blue diamonds, respectively.
We note that both methods find the pion and the first excited state, even with similar error bars.
The variational method with an expanded interpolator basis is numerically more robust than the $\cosh$ method and therefore able to go two time steps further to the right.} \label{fig:varmethcosh33}
\end{figure}

The Generalized Pencil-of-Function method \cite{Aubin:2010jc} or, more catchy, the Time-Shift Trick allows one to obtain additional interpolators ``for free''. 
As the name suggests, new interpolators are constructed by time-shifting other interpolators.
I will label the time-shift operator $T (\delta t) = T^\dagger (\delta t)$, with the amount of time shift $\delta t$.
Its action on an operator $\mathcal{O} (t)$ is defined by $T(\delta t) \mathcal{O} (t) = \mathcal{O} (t - \delta t)$ and due to time invariance one sees that
\begin{align}
\braket{ \mathcal{O}_2 (t_2) T (\delta t) \mathcal{O}_1 (t_1) } =&\, 
\braket{ \mathcal{O}_2 (t_2) \mathcal{O}_1 (t_1 - \delta t) } \nonumber \\
=&\, \braket{ \mathcal{O}_2 (t_2 + \delta t) \mathcal{O}_1 (t_1) } \nonumber
\end{align}
and therefore 
\begin{equation} \label{eq:ttotheleft}
\mathcal{O} (t) T (\delta t) = \mathcal{O} (t + \delta t).
\end{equation}
Using this, one can define a new interpolator $\intp'_i \equiv T(\delta t) \intp_i$ and sees immediately that
\begin{align}
\braket{ s_i | \intp'_j | 0 } =&\, \braket{ s_i | T(\delta t) \intp_j | 0 } = \exp(- E_i \delta t) \braket{ s_i | \intp_j | 0 } \nonumber \\
=&\, \exp(- E_i \delta t) a_{ij}. \nonumber
\end{align}
Therefore, $\intp'_i$ is, in general, a new linearly independent interpolator. 

\paragraph{Proof of principle.---}

In this final paragraph, I show that the variational method with an expanded interpolator basis and the Time-Shift Trick actually work with ``real data''. 
I investigate the pion with the two interpolators $\Pi$ and $A_0$ and the time-shifted interpolator $\Pi\ T(\delta t)$ so that I can account for the pion, the excited pion and the backward running pion.
Using Eq.~\eqref{eq:ttotheleft}, one obtains the correlation matrix
\begin{widetext}
\[
C(t) = \left( \begin{array}{r@{|}lr@{|}lr@{|}l} 
\braket{ \Pi (t) & \Pi^\dagger (0) } & \braket{ \Pi (t) & A_0^\dagger (0) } & \braket{ \Pi (t + \delta t) & \Pi^\dagger (0) } \\
\braket{ A_0 (t) & \Pi^\dagger (0) } & \braket{ A_0 (t) & A_0^\dagger (0) } & \braket{ A_0 (t + \delta t) & \Pi^\dagger (0) } \\
\braket{ \Pi (t + \delta t) & \Pi^\dagger (0) } & \braket{ \Pi (t + \delta t) & A_0^\dagger (0) } & \braket{ \Pi (t + 2 \delta t) & \Pi^\dagger (0) } \\
\end{array} \right).
\]
\end{widetext}
I use this matrix in the standard variational method and find the eigenvalues of $G(t_0, t_1)$ in Eq.~\eqref{eq:varmatrix}.
To obtain optimal results, I vary $t_0, t_1$ and $\delta t$.
It turns out that for this example, $t_1 - t_0 = 2$ and $\delta t = 14$ are good choices.
For the sake of completeness, I compare these results to the results obtained with the $\cosh$ method, where I also kept $t_1 - t_0 = 2$.
The results are shown in Figure~\ref{fig:varmethcosh33}.
Both methods find perfect plateaux for the pion (the variational method with an expanded interpolator basis also for the backward running pion) and reasonably nice plateaux for the excited pion.
The error bars are similar and the eigenvalues agree within the error bars in the region of the plateau.
So while the results are comparable, I prefer to use the variational method with an expanded interpolator basis over the $\cosh$ method because of its numerical stability and robustness (which also allows the former method to reach higher $t_0$ than the $\cosh$ method).
Furthermore, the variational method with an expanded interpolator basis is faster than the $\cosh$ method, although this is usually not relevant with today's computers.

\begin{figure}
  \begin{picture}(1,0.63941617)%
    \includegraphics[width=\unitlength]{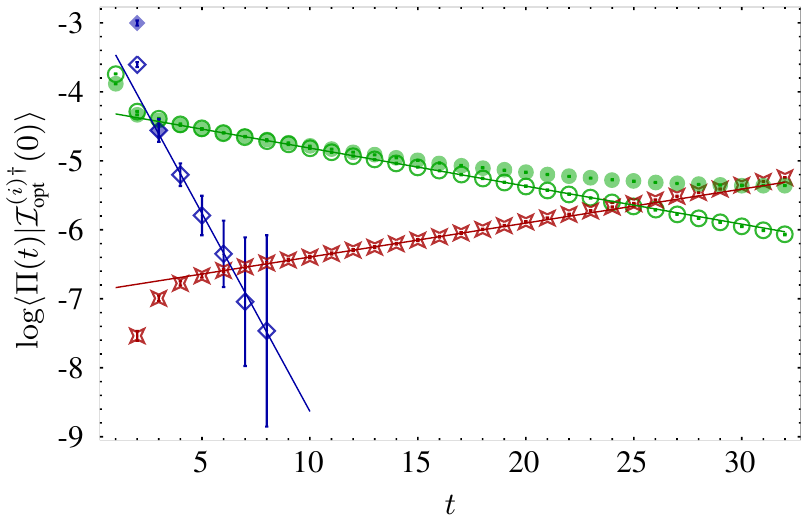}%
  \end{picture}%
\caption{The correlators of the optimal interpolators with the $\Pi$ interpolator, based on the $\cosh$ method (filled symbols) and the AIM (open symbols). 
The lines are best linear fit lines to the $\log$ of the correlators obtained with the variational method with the expanded interpolator basis in the region where the excited states have decayed sufficiently.
The backward running pion (red stars) is by construction only found with the variational method with an expanded interpolator basis.
The forward running pion (green circles) is influenced by the backward running state for the $\cosh$ method while it is a perfect exponential up to large times for the variational method with an expanded interpolator basis.
Also, the excited pion state (blue diamonds) is resolved better with the variational method with an expanded interpolator basis than with the $\cosh$ method.} \label{fig:optcorr}
\end{figure}

The big advantage of the variational method with an expanded interpolator basis over the $\cosh$ method, however, are the optimal interpolators.
For each method, I construct the optimal interpolators as described above and show the correlators $\braket{ \Pi (t) | \intp^{(i) \dagger}_\text{opt} (0)}$ in Figure~\ref{fig:optcorr}.
The variational method with an expanded interpolator basis yields an optimal pion, i.e., one that has an exponential (linear in the logarithmic plot) behavior throughout the plot range.
The backward running pion is also very good and the excited pion is reasonable.
The $\cosh$ method, on the other hand, yields a poor excited state and a pion which is usable up to $t \approx 10$.
For larger $t$, the backward running state starts to strongly influence the signal.

In conclusion, the variational method with an expanded interpolator basis has proven to be a robust method that is at least as good as the $\cosh$ method in finding the energy eigenvalues but has its real strengths in the construction of optimal interpolators.
The Time-Shift Trick is also very useful and can be applied wherever additional interpolators are needed, in many cases even at no extra computational cost.

\paragraph{Acknowledgments.---}

I would like to thank my colleagues of the Regensburg group (RQCD) for letting me use the lattice raw data for the proof of principle.
This work has been supported in part by the Deutsche Forschungsgemeinschaft (SFB/TR 55). 
The computations were performed on the QPACE systems of the SFB/TR 55 and the SuperMUC system at the LRZ/Germany using the Chroma software system \cite{Edwards:2004sx} and the BQCD software \cite{Nakamura:2010qh}  including improved inverters \cite{Nobile:2010zz,LuscherOpenQCD}.

\bibliography{varmeth}

\end{document}